\documentclass[twocolumn,aps,pre,floatfix,superscriptaddress,showpacs]{revtex4}
\usepackage{graphicx} \usepackage{bm} \usepackage{epsfig}

\begin{document}

\title{Glassy behaviour in an exactly solved spin system with a 
ferromagnetic transition}

\author{Robert L. Jack} \affiliation{Rudolf Peierls Centre for
Theoretical Physics, University of Oxford, 1 Keble Road, Oxford, OX1
3NP, UK}
 
\author{Juan P. Garrahan} \affiliation{School of Physics and
Astronomy, University of Nottingham, Nottingham, NG7 2RD, UK}
 
\author{David Sherrington} \affiliation{Rudolf Peierls Centre for
Theoretical Physics, University of Oxford, 1 Keble Road, Oxford, OX1
3NP, UK}
 
\begin{abstract}
We show that applying simple dynamical rules to Baxter's eight-vertex
model leads to a system which resembles a glass-forming liquid. There
are analogies with liquid, supercooled liquid, glassy and crystalline
states. The disordered phases exhibit strong dynamical heterogeneity
at low temperatures, which may be described in terms of an emergent
mobility field.  Their dynamics are well-described by a simple model
with trivial thermodynamics, but an emergent kinetic constraint.  We
show that the (second order) thermodynamic transition to the ordered
phase may be interpreted in terms of confinement of the excitations in
the mobility field.  We also describe the aging of disordered states
towards the ordered phase, in terms of simple rate equations.
\end{abstract}
\pacs{64.60.-i,64.70.Pf,05.50.+q}
\maketitle

Despite many years of study of glass-forming liquids, the most
suitable paradigm in which to discuss the `glass transition' and its
associated dynamical phenomena remains controversial. In recent years,
there has been progress \cite{GCTheory,Berthier2003,Whitelam2004},
driven by the idea that the dynamical properties of glass-formers may
be characterised by a zero temperature dynamical fixed
point\cite{Whitelam2004}.  This is in contrast to the predictions of
other theories which involve a finite temperature singularity in the
dynamics \cite{MCT,Biroli-Bouchaud} or the thermodynamics of the
relevant system\cite{Tarjus-Kivelson,Xia-Wolynes,Mezard-Parisi}.

However, there remains an important qualitative difference between
physical glass-formers and the models, such as the
Fredrickson-Andersen (FA)\cite{FAModel} or East\cite{EastModel}
models (see \cite{Ritort-Sollich} for a review) studied
in~\cite{GCTheory,Berthier2003,Whitelam2004}: the crystal phase is
completely absent from these models.  As a result, they are relevant
if the glass-forming liquid is in a long-lived (but metastable)
supercooled phase. Further, interplay between the thermodynamic
singularity associated with the transition to the crystal and the
dynamical fixed point associated with the glass is certainly possible,
but the FA and East models cannot capture these phenomena since their
thermodynamics are trivial.  There has been recent
work\cite{Cavagna2003, Franz2001} (see also \cite{Swift2000})
investigating these issues
with regard to glassy systems, although without reference to a glassy
fixed point at zero temperature.

The extent to which the FA and East models can be viewed as pictures
of real glasses is therefore contingent on two main assumptions.
Firstly, the behaviour of the supercooled state should not be affected
by the proximity of the freezing transition, since the FA and East
models regard glassy slowing down as a purely dynamical effect, not
requiring a thermodynamic transition.  Secondly, the `mobility field'
represented by the spins in these models should emerge naturally from
atomistic degrees of freedom of the glass-former.

One class of models in which this latter effect is demonstrated are
the two-dimensional plaquette models
\cite{Lipowski,Garrahan-Newman,barca}, in which an effective dynamical
constraint emerges naturally from a simple spin model.  There are free
excitations in the spin field which are naturally interpreted as a
mobility field.  At temperatures lower than the glassy onset
temperature, $T_o$ \cite{Brumer,Berthier2003}, the dynamics are
strongly heterogeneous, and slow down rapidly with decreasing
temperature.

In this paper, we address the other issue mentioned above: how are
dynamics of metastable supercooled states affected by the presence of
the freezing transition?  We study dynamics in the eight-vertex model,
whose thermodynamics were solved by Baxter\cite{BaxterBook}.  The
model may be treated as a generalisation of the square plaquette model
\cite{Lipowski,barca}; the effect of this generalisation is to to
introduce a (second order) phase transition to an ordered state at a
finite temperature $T_c$. We identify this with a freezing transition,
and investigate the dynamics around this transition. The transition
temperature $T_c$ may be varied with respect to the glassy onset
temperature $T_o$. This fact, together with the exactly solved
thermodynamics, gives an extra degree of control to our simulations.

We will show that we may prepare long-lived `supercooled' states below
$T_c$, whose dynamics are controlled by the effective dynamical
constraint of the plaquette model, and are not affected by the
freezing transition for times shorter than the lifetime of the
supercooled state.

To be more precise, the dynamics of the system within the supercooled
state resemble those of strong glasses, and arise from diffusing
point-like excitations in a mobility field
\cite{GCTheory,Whitelam2004,Berthier2003}. Considering supercooled
states at different temperatures, we find that they obey dynamical
scaling consistent with a zero temperature fixed point. The presence
of the ferromagnetic transition means that these states have a finite
lifetime, but it does not affect the dynamics on timescales shorter
than this lifetime.  This is consistent with the assumptions made when
modelling glass-formers using models without an ordering transition
\cite{GCTheory,Whitelam2004,Berthier2003}.

The form of the paper is as follows: in section~\ref{sec:model}, we
describe the model, identify relevant energy scales and their
hierarchy, and discuss the relation between the spins of our model and
the atomistic degrees of freedom of a physical glass former.  We
discuss the nature of the ordered and disordered phases of the model
system in section~\ref{sec:static}: we then use this information to
interpret simulations of the dynamics of the model in
section~\ref{sec:dynamic}. Finally, we summarise our results in
section~\ref{sec:conc}, and discuss their significance for models of
the glass transition.

\section{The model}
\label{sec:model}

The zero-field eight vertex model, solved by Baxter\cite{BaxterBook} 
may be expressed 
either in terms of its original vertices, or as an Ising model with 
Hamiltonian:
\begin{eqnarray}
H & = \sum_{ij} & \left[ 
 - D \sigma_{ij} \sigma_{i+1,j} \sigma_{i,j+1} \sigma_{i+1,j+1}
\right. \nonumber \\ & & \left.
 - J  \left( \sigma_{ij} \sigma_{i+1,j+1} + \sigma_{i+1,j} \sigma_{i,j+1} 
      \right)  \right]
\label{equ:H_spins}
\end{eqnarray}
where the $\{\sigma_{ij}\}$ are Ising spins on a square lattice. We
note that the Ising coupling is between \emph{next nearest} neighbours
on the square lattice: at $D=0$ there are two independent sublattices,
with Ising coupling, $J$ within each sublattice. There is a transition
at $J = (T/2) \sinh^{-1}(1)$ to a fourfold degenerate ordered state
(there are two ferromagnetic and two antiferromagnetic ground states,
related by flipping all the spins on either sublattice). As $D$ is
increased from zero, the lattices become coupled, and the transition
moves to a higher temperature: the critical temperature $T_c$
satisfies:
\begin{equation}
\sinh(2J/T_c) = \exp(-2D/T_c)
\label{equ:def_tc}
\end{equation}
The transition to an ordered state occurs for all finite $J$: we
also note that if $D>J$ then the transition temperature will be much
larger than $J$. 

Thus far we have considered only static (thermodynamic) properties of the
eight-vertex model. In order to study the time evolution of the this
model, we must specify dynamical rules. We use simple spin flips with
rates given by Glauber dynamics. We refer to the combination of the 
Hamiltonian and the dynamical rules as the spin-flip eight vertex (SEV)
model.

If we set $J=0$ in the SEV model, we arrive at the (two dimensional)
plaquette model\cite{Lipowski,Buhot2002,Espriu2003cm}.  In this limit
there is no ordering at any finite temperature: all two point static
correlations vanish. This may be most easily demonstrated by noting
that if $J=0$ then the Hamiltonian is invariant under flipping all of
the spins in any row or column of the square lattice. The dynamics of
this model are dominated by a zero temperature dynamical fixed point
for temperatures $T$ that are small compared to $D$. Since we are
studying slow dynamics we work throughout at $T<D$.

We have now identified two temperature scales in the problem: the
onset of glassy dynamics occurs at $T=T_o\sim D$ and the critical
point in the system is at $T=T_c$. If $T_c>T_o$ then we expect the
slow dynamics to be observable only in the ordered phase: the more
interesting case is $T_o>T_c$, in which case the dynamics are
slow near the transition, and we may investigate the effect of
the effective kinetic constraint as we cool the system through $T_c$.
We therefore work at $T_c < D$: from (\ref{equ:def_tc}), this means that
$J<T_c$. As a result we have the hierarchy $J < (T,T_c) < D$ which is
obeyed throughout this paper.

\subsection{Relation of this work to physical glass formers}
\label{sec:glass}

Before investigating the SEV model more closely, we establish the
relationship between this model and physical glass-formers: it is not
obvious at first sight precisely 
how a model of Ising spins should be related to
a atomistic system.  The spin variables represent the microscopic
degrees of freedom of the glass-former. This is distinct from the more
heuristic approach taken in the FA and East models in which the spins
represent a coarse-grained `mobility field'. The SEV model is more
similar to the plaquette model\cite{barca} in that the effective
kinetic constraint responsible for the critical slowing down at zero
temperature is not inserted explicitly, but arises from the
combination of the Hamiltonian and simple spin-flip dynamics. In
section~\ref{sec:dynamic} we will comment briefly on how the dynamics
may be interpreted within a `mobility field' picture like that of the
FA model.

In the previous section we identified the two temperature scales in
the model as the glassy onset temperature $T_o\sim D$ and the
transition temperature $T_c<T_o$.  These separate the behaviour of the
system into three regimes.  We argue that the high temperature phase
of the SEV model with $T>(T_o,T_c)$ resembles a liquid-like state of
the atomistic system, since it lacks any two point correlations
between the spins.

The second regime is $T_c<T<T_o$: there are still no static
correlations between the spins but there are strong \emph{dynamical}
correlations. This state resembles a viscous liquid whose relaxation
time is large compared to microscopic timescales.  We emphasise that
the crossover between this regime and the high temperature behaviour
is smooth: there is no sharp transition at $T_o$. In the viscous
liquid, the atomistic degrees of freedom are `jammed' over large
regions of the system: relaxation in these regions is frustrated by
large energy barriers.  Dynamical heterogeneity then arises naturally,
due to the presence of mobile regions where the energy barriers are
smaller than average.  There are very many paramagnetic states in the
spin system, even at low temperatures (compared to $D$): these
resemble the many possible jammed states of the glass former.

Having argued that the paramagnetic phase of our model is liquid-like,
and shows glassy behaviour at low temperatures, it is natural to
interpret the transition in the model as a a freezing transition. We
identify the ferromagnetic phase with the crystalline states of the
glass-former.  As the temperature is lowered through $T_c$ the entropy
falls rapidly as the very many paramagnetic (jammed) states are now
thermodynamically unstable with respect to the ferromagnet.  The
effect of the dynamical fixed point on the transition between
paramagnet and ferromagnet is the main subject of the following
sections. In particular we show that `supercooling' of the
paramagnetic state is possible as long as $T_c \ll D$.

\section{Static properties of the SEV model}
\label{sec:static}

\begin{figure} \begin{center}
\epsfig{file=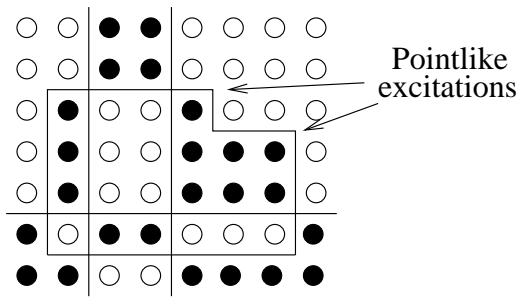,width=0.8\columnwidth} 
\end{center}
\caption{Sketch showing the relation between spin states, excited
plaquettes and domain walls. Up (down) spins are represented by filled
(empty) circles.  Domain walls form closed loops with excited
plaquettes (energy cost $2D$) at each isolated corner (there is no
energy cost associated with wall crossings). A domain wall costs
energy $2J$ per unit length when running though a ferromagnetically
ordered region. Interactions between domain walls reduce this tension
in disordered states.}
\label{fig:poly}
\end{figure}

We now discuss the microscopic structure of the ferromagnetic and 
paramagnetic states in the model of equation (\ref{equ:H_spins}). 
This will allow us to identify the relevant correlation functions for
our study of the dynamics of the system.  We describe the
paramagnetic state in terms of small deviations from the behaviour of
the model with $J=0$, and the
ferromagnetic state in terms of excitations in an ordered background. This
will lead us to interpret the transition in terms of free defects above
$T_c$ that become confined at the transition, forming composite
excitations. We will also show that these descriptions are valid even
rather close to the transition, despite being based on expansions around
the fully ordered or fully disordered states. In other words, the
critical region is very narrow.

We begin by recalling some results for the
$J=0$ limit of the model\cite{Buhot2002,Espriu2003cm}. At $J=0$, 
we write $p_i=\sigma_{ij} \sigma_{i,j+1} \sigma_{i+1,j} \sigma_{i+1,j+1}$, 
and the Hamiltonian reduces to 
\begin{equation}
H_{J=0} = - D \sum_i p_i
\label{equ:H_plaq}
\end{equation}
where the $p_i$ are Ising-like degrees of freedom, defined on the 
plaquettes of the square lattice. In the thermodynamic limit, these
plaquettes are independent degrees of freedom, that define the state
of the spin system, up to transformations that flip all the spins in 
any row or column (leaving $H_{J=0}$ invariant). We observe that this 
results in a ground state entropy proportional to the linear size of the
system, $L$ (there are $N=L^2$ spins in the system). 

In finite systems,
the presence of boundary conditions may impose constraints on the 
plaquettes. For example, imposing periodic
boundaries on the spins means that the number of excited plaquettes in
all rows and columns must be even (we believe that this fact led to the 
strong finite size effects seen in \cite{Espriu2003cm}). 

As discussed in~\cite{Espriu2003cm}, 
the low temperature states of the model with $J=0$
are best interpreted in terms of closed loops
with excited plaquettes at each vertex (see figure~\ref{fig:poly}). 
If we move across the lattice,
any spin flip is accompanied by our crossing the perimeter of a loop. 
If $J=0$ then each plaquette is independent: the vertices of the loops
are a free lattice gas with density $(1+e^{2D/T})^{-1}$. The free
energy per site is simply
\begin{equation}
f_{J=0} = -T \log [2\cosh(D/T)]
\label{equ:f_J=0}
\end{equation}

At finite $J$, we make use of Baxter's solution of the eight vertex 
model\cite{BaxterBook}. In appendix~\ref{app:series}, 
we show that the free energy per site for $T\gg T_c$ is given 
approximately by 
\begin{equation}
f_\mathrm{pm}^{(0)} = - T\log \left[ e^{D/T} \cosh(2J/T) + e^{-D/T}
\right]
\label{equ:f_pm}
\end{equation}
Since we work exclusively
at $J<T$, equation~(\ref{equ:f_pm}) is rather close to the $J=0$ 
expression~(\ref{equ:f_J=0}).

We focus on two correlation functions,
the density of excited plaquettes $n_p$, and the density 
of broken Ising bonds, $n_b$. 
Their definitions are:
\begin{eqnarray}
n_p & = & \frac{1}{2}(1-\langle \sigma_{ij} \sigma_{i,j+1} \sigma_{i+1,j} 
   \sigma_{i+1,j+1}\rangle) 
\\
n_b & = &\frac{1}{4}(2-
\langle \sigma_{ij} \sigma_{i+1,j+1} + \sigma_{i+1,j} \sigma_{i,j+1}
\rangle )
\end{eqnarray}

\begin{figure} \begin{center}
\epsfig{file=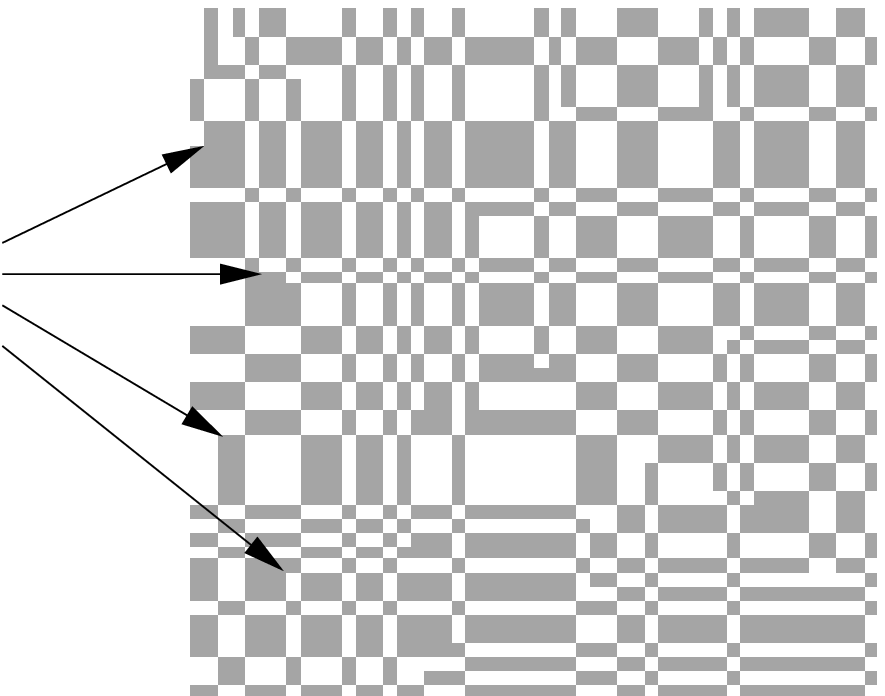,width=0.49\columnwidth}
\epsfig{file=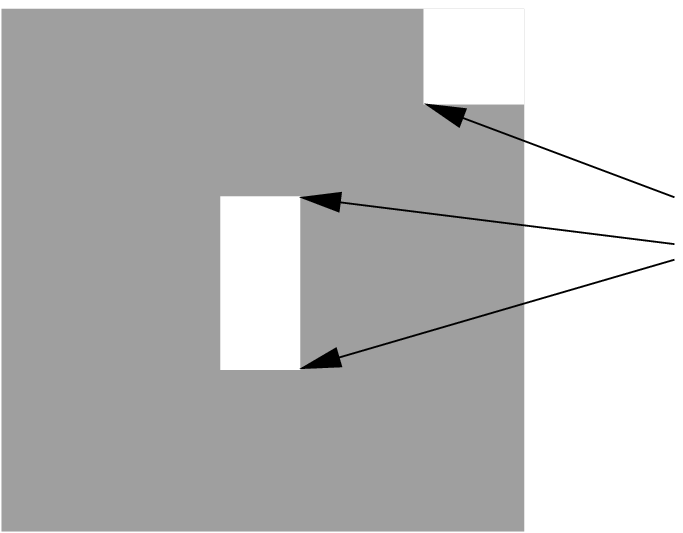,width=0.49\columnwidth}
\end{center}
\caption{Typical states of the spin system above $T_c$ (left)
and below $T_c$ (right). Some of the excited plaquettes are identified
by arrows. They are `free' above $T_c$, but `confined' on corners of
rectangular excitations in the ferromagnetic phase.}
\label{fig:states}
\end{figure}

In the representation of figure~\ref{fig:poly}, the concentration of
vertices is given by $n_p$. The parameter $n_b$ is related to the
total perimeter of the closed loops of that figure, and measures the
spatial ordering of the vertices.  The free plaquettes observed at
$J=0$ have $n_b=(1/2)$.  As $n_b$ is reduced, the reduction in the
loop perimeter starts to constrain the positions of the vertices, and
spatial correlations appear.

The internal energy per site is given by 
\begin{equation}
\langle H/N \rangle = 2Dn_p + 4Jn_b - (D + 2J)
\end{equation}
so we have $n_p = (1+\partial f/\partial D)/2$ and $n_b = (2+\partial
f/\partial J)/4$. Using (\ref{equ:f_pm}), the paramagnetic state has
\begin{eqnarray}
n_p & \simeq & [1+e^{2D/T}\cosh(2J/T)]^{-1}
\\
n_b & \simeq & (1/2) - \frac{\sinh(2J/T)}{\cosh(2J/T)+e^{-2D/T}}
\end{eqnarray}
We see that introducing $J$ leads to small negative corrections to the
$J=0$ values of $n_p$ and $n_b$. However, for $T<D$, the concentration
of excited plaquettes is still approximately $c=e^{-2D/T}$ and these
plaquettes are only weakly interacting since $n_b\simeq(1/2)$.  A
typical configuration is shown in figure~\ref{fig:states}: there are
no two point correlations between the spins, but the loop vertices are
dilute since $T\ll D$.

We now consider the system at temperatures lower than $T_c$.  As shown
in appendix~\ref{app:series}, a good approximation to the free energy
per site in the ferromagnetic phase is
\begin{equation}
f_\mathrm{fm}^{(0)} = -D - 2J - \frac{T e^{-8D/T}}{16\sinh^2(2J/T)}
\label{equ:f_fm}
\end{equation}
which is valid for $J<T<T_c<D$. In this regime
\begin{eqnarray}
n_p & = & \frac{e^{-8D/T}}{4 \sinh^2 (2J/T)}  \label{equ:nbfm} \\
n_b & = & \frac{e^{-8D/T}}{16 \sinh^2(2J/T) \tanh(2J/T)}
\label{equ:npfm}
\end{eqnarray}

For excitations in a ferromagnetic background, $n_b$ is directly
related to the perimeter of the closed loops shown in
figure~\ref{fig:poly}.  Equations~(\ref{equ:nbfm})
and~(\ref{equ:npfm}) are therefore consistent with rectangular
excitation loops with an excited plaquette at each corner (total
energy cost $8D$). The expectation of the loop perimeter is
approximately $(T/2J)$.  This is much smaller than the typical spacing
between loops, given by $(8J/T)e^{4D/T}$.  This situation is sketched
in figure~\ref{fig:states}.

The density of loops is given by the number of ways of forming such a
loop, multiplied by a Boltzmann factor, $e^{-8D/T}$. We therefore
identify the entropy per loop as $S_\mathrm{loop}\sim -2\log
[4\sinh(2J/T)]$.  The apparent divergence of the perimeter (and
therefore the entropy) at small $J$ represents the breakdown of the
ordered state which happens at $T_c$.  The transition to the
paramagnetic state occurs when the energy cost for adding two vertices
to the loop ($4D$) is balanced by the entropy gain associated with
adding an extra rectangular segment to the loop.  This entropy gain is
approximately $S_\mathrm{loop}$. We may therefore obtain an estimate
of the transition temperature by setting $4D/T_c =
S_\mathrm{loop}$. The result is
\begin{equation}
4 \sinh(2J/T_c) \simeq \exp(-2D/T_c)
\end{equation}
which differs from the exact result (\ref{equ:def_tc}) only by the
constant leading factor of 4. The real transition temperature is lower
than that predicted by this method because interactions between the
loops act to reduce the energies at large perimeters.

Thus we interpret the transition to the paramagnet as deconfinement of
localised composite excitations. The state becomes disordered when the
loop size becomes comparable with the spacing between loops.

The magnetisation and correlation lengths may also be calculated in
similar series (see appendix~\ref{app:series}).  The magnetisation,
$M_0$, may be used to calculate the fraction of spins opposed to the
mean spin,
\begin{equation}
n_s = \frac{1-M_0}{2} = \frac{e^{-8D/T}}{256 \sinh^4(2J/T)}
\end{equation}
Assuming that the lowest-lying excitations are rectangular domains
with four excited plaquettes per domain, we expect a relation of the
form $n_s = (n_p/4)(n_b/n_p)^2$. We see that this is true for small
$J/T$ [recall that $(J/T)$ is a small number, although expansions
about $J=0$ are not valid in general since we are in an ordered
phase].

Thus we have arrived at the following picture of the thermodynamics of
the SEV model. There is a density of excited plaquettes $n_p$, which
sit on the corners of overlapping closed loops. The total loop
perimeter is measured in terms of the parameter $n_b$: the point
$n_b=(1/2)$ is the maximally disordered spin field, in which the
excited plaquettes are free. Away from the critical region (which is
narrow), the excited plaquettes are nearly free in the paramagnetic
phase: in the ferromagnetic phase then they are confined on corners of
rectangular loops whose typical size is much smaller than their
inverse density.

So far, our microscopic arguments have been purely thermodynamic: we
have not considered any dynamics. In the next section, we investigate
the dynamics of the SEV model.
 
\section{Dynamics}
\label{sec:dynamic}

This section contains the key results of this paper. We briefly describe the
dynamics of the paramagnetic state, which are essentially independent of $J$.
We then discuss the onset of ordering as the temperature is lowered through
$T_c$. We will find that supercooled states exist near $T_c$, which are 
well described by a simple `mobility field' picture for times shorter
than their lifetime. We then discuss the extent to which these states can 
be regarded as metastable, and what determines their lifetimes. 

We begin with a very brief review of the dynamics in the
paramagnetic state with $T<T_o$. 
Since $J$ may be treated perturbatively in this
regime, we write the Hamiltonian as in equation~(\ref{equ:H_plaq}).
Flipping a single spin involves flipping of the four plaquette variables
adjacent to that spin. Thus spins adjacent to four unexcited plaquettes
have a flipping rate that is suppressed by a factor $e^{-8D/T}$. However,
the flip rate of spins
adjacent to exactly one excited plaquette are suppressed only by a
factor $e^{-4D/T}$. The
excited plaquettes mark mobile regions in which spin flips are rather
likely. Thus the model resembles kinetically constrained systems such as
the FA model\cite{FAModel}.

The relaxation time of the spins depends on the temperature and
on the density of excited plaquettes, according to 
$\tau \sim n_p^{-1} e^{4D/T}$.
This arises from localised one-dimensional diffusion of pairs
of excited plaquettes\cite{Buhot2002,Espriu2003cm}. In equilibrium,
we have $n_p\simeq c=e^{-2D/T}$, so the relaxation time diverges as $c^{-3}$. 
More precisely, we have the scaling relation\cite{Buhot2002}
\begin{equation}
\langle \sigma_{ij}(t_w) \sigma_{ij}(t+t_w) \rangle_{c,\mathrm{eq}}
 = f(c^3 t)
\label{equ:tau_scaling}
\end{equation}
for the on-site autocorrelation function in equilibrium at a given
value of $c$.  This is strong glass behaviour in the classification of
Angell \cite{Angell}.

We now turn to results for the SEV model below $T_c$, where $J$ may
not be treated perturbatively.  We discuss the phenomenological
similarities and differences between this model and physical glass
formers. We then interpret this behaviour with the aid of mean field
rate equations.

\subsection{Existence of supercooled states}
\label{sec:sc_sub}
\begin{figure} \begin{center}
\epsfig{file=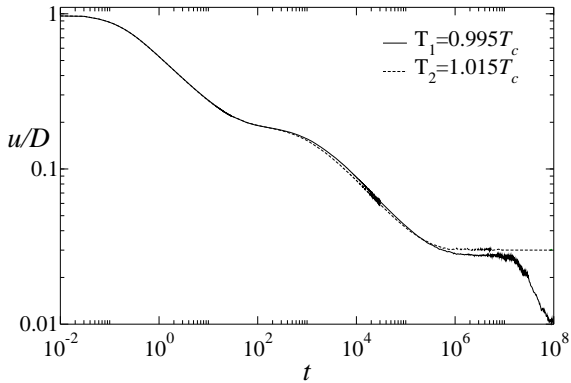,width=0.98\columnwidth}
\end{center}
\caption{Behaviour of the system after two quenches from infinite 
temperature to just below and just
above $T_c = 0.45 D$. The two plateaux correspond to the onset of 
activated dynamics, and equilibration in the disordered (glassy) state. 
However, the disordered state is unstable below $T_c$, as is clear at longer
times. }
\label{fig:energy_q}
\end{figure}

We start this section by demonstrating how a supercooled state may be
formed below $T_c$.  We investigate the dynamics of the system
by means of simulations which use a continuous time Monte
Carlo algorithm\cite{Newman-Barkema}, with periodic boundary
conditions. 
The number of excited plaquettes in each row and column is
constrained to be even in this treatment, so the linear size of the
system must be greater than $(2/n_p)$ for reliable results.  
Some of the main features of the dynamics of the
SEV model are shown in figure~\ref{fig:energy_q}.  We
measure the energy density of the system with respect to the ground
state:
\begin{equation}
u = \langle H/N \rangle + D + 2J
\end{equation} 
The dashed trace in figure~\ref{fig:energy_q} shows the internal
energy density after quenching to a temperature, $T_1$, such that
$T_c<T_1<D$.  After an initial transient, the system ages in a power
law fashion towards equilibration at $u \sim 2D e^{-2D/T_1} +2J$. The
plateau at $(u/D)\simeq0.2$ is a characteristic feature of models with
kinetic constraints (whether explicit\cite{FAModel} or
emergent\cite{Garrahan-Newman,Davison}).  It represents the onset of
`activated' dynamics. The equilibration time for the paramagnet scales
as a power of $c=e^{-2D/T}$.  All these features are seen in the model
with $J=0$ \cite{Lipowski,barca}.

In contrast, the full trace in figure~\ref{fig:energy_q} shows the
behaviour on quenching to a temperature $T_2$ satisfying $T_2<T_c<D$,
but with $T_2$ close to $T_c$. The behaviour resembles that of the
quench to $T_1$, including apparent equilibration at $u\sim
2De^{-2D/T_2} +2J$. However, this is state is not stable, and the
energy falls further at longer times. This behaviour resembles that of
glass-formers, where the state on the lower plateau would be called a
supercooled liquid. The behaviour is also qualitatively similar to
that observed in \cite{Cavagna2003}.

\begin{figure} \begin{center}
\epsfig{file=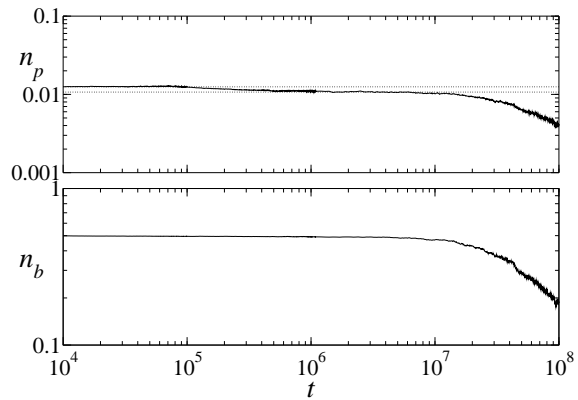,width=0.98\columnwidth}
\end{center}
\caption{Behaviour of the SEV model after a quench from $T_1=1.02 T_c$
to $T_2=0.996 T_c$ with $T_c=0.45 D$. We show concentrations of
excited plaquettes, $n_p$, and broken bonds, $n_b$. The dotted lines
mark $c_{1,2}=e^{-2D/T_{1,2}}$. }
\label{fig:np_cool}
\end{figure}

In order to focus on the supercooled states, we show further
simulations in figure~\ref{fig:np_cool}. The system is cooled through
$T_c$ in a single small step. We plot $n_p$ and $n_b$ as a function of
time.  From the plot of $n_p$, we see that the density of free
excitations responds relatively quickly to the change in temperature:
it falls from $c_1=e^{-2D/T_1}$ to $c_2=e^{-2D/T_2}$, where it appears
to stabilise.

The response of $n_b$ to the cooling is much slower. Recall that this
correlation function measures the clustering of excited
plaquettes. This clustering is a much slower process than the creation
and annihilation steps leading to a change in the concentration of
excitations.  Looking at the late times in figure~\ref{fig:np_cool}
when the clustering does start to occur, the system ages towards the
ferromagnetic state with both $n_p$ and $n_b$ falling together. Taking
the two traces in figure~\ref{fig:np_cool} together, we see that there
are two separate timescales: one is associated with changes in $n_p$;
the other with changes in $n_b$.

Turning to the supercooled state itself, it is clear from
figure~\ref{fig:np_cool} that it has $n_p\simeq c = e^{-2D/T}$ and
$n_b \simeq (1/2)$. This resembles closely the state that would be
formed if we set $J=0$ in the Hamiltonian. Thus the effect of the
interactions between plaquettes (the term proportional to $J$ in the
Hamiltonian) is to set the lifetime of the supercooled state. The
properties of the state itself are independent of $J$.  We conclude
that the supercooled state in the SEV model can be well described by
the much simpler plaquette model of equation~(\ref{equ:H_plaq}): a
kinetically frustrated model with trivial thermodynamic properties.
This is the assumption made when describing glass-formers by simple
models of dynamical heterogeneity\cite{GCTheory,Berthier2003}: in the
SEV model, this assumption seems to be reasonable.

\begin{figure} \begin{center}
\epsfig{file=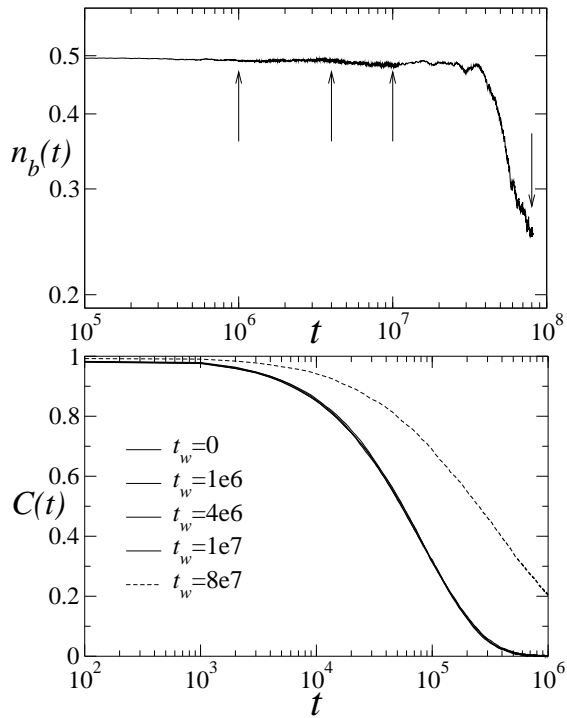,width=0.9\columnwidth}
\end{center}
\caption{(Top) fraction of broken bonds against waiting time in the
supercooled state: $T=0.996 T_c$, $T_c=0.444D$.  (Bottom)
Autocorrelation function $C(t)=\langle \sigma_{ij}(t+t_w)
\sigma_{ij}(t_w)\rangle$ for the waiting times marked in the top
figure. The two-time function appears stationary until the system
starts to nucleate at around $3\times 10^7$ Monte Carlo
sweeps.  } \label{fig:stat}
\end{figure}

A key property of supercooled states is that two-time correlation
functions should be stationary within the supercooled state. That is,
expectation values of the form $\langle A(t_w) A(t+t_w) \rangle$
should be independent of $t_w$ as long as $(t+t_w)$ is less than the
lifetime of the states. In figure~\ref{fig:stat}, we show that the
supercooled state with $n_p=c$ and $n_b=(1/2)$ has this
property. Since the excited plaquettes are uncorrelated in this state,
it may be prepared manually, without the need for simulation. Further,
the single spin autocorrelation function in the supercooled state is
the same as that in the model with $J=0$ at the same temperature. It
therefore obeys the dynamical scaling law of
equation~(\ref{equ:tau_scaling}), and is independent of $J$.

As a final comment on figure~\ref{fig:stat}, we note that 
the criterion that the supercooled state should be long-lived is
fulfilled, since stationarity holds over timescales much longer than the 
single-spin relaxation time. Thus the data in that figure is consistent
with the picture of a supercooled state that appears to equilibrate
in a metastable basin.

\subsection{Lifetime of supercooled states}

\begin{figure} \begin{center}
\epsfig{file=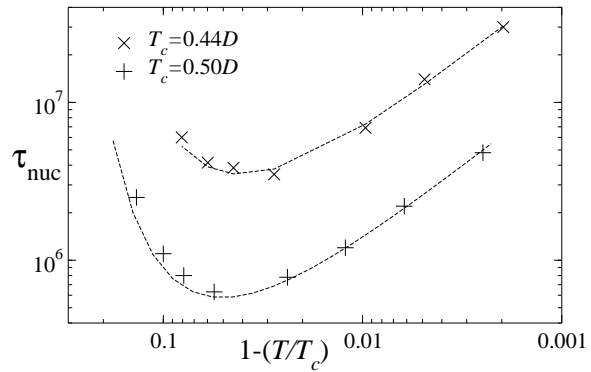,width=0.98\columnwidth}
\end{center}
\caption{Nucleation times at
$T_c=0.50D$ and $T_c=0.44D$, measured by averaging over several runs
similar to those in figure~\ref{fig:stat}.  The temperature increases
from left to right, but note that the zero of the logarithmic scale is
at $T_c$, away to the right. The dashed lines are fits to the form of
equation (\ref{equ:t_nuc}) using the same value of $\gamma=0.17$. This
form captures the qualitative features of the curve: a power law
increase near $T_c$, and an exponential divergence at low
temperatures.  }
\label{fig:nuc}
\end{figure}

Having identified a supercooled state at temperatures near $T_c$, we
now proceed to discuss its lifetime. We prepared states with $n_p=c$
and $n_b=(1/2)$ and measured the time taken for $n_b$ to fall to
$0.45$. Results are shown in figure~\ref{fig:nuc}.  The lifetime gets
very large both near $T_c$, and in the limit $T\to0$.

Near $T_c$, the states are supercooled.  At very low temperatures, the
lifetimes are similarly long, but in this case they are of the same
order as the spin relaxation time. These states are not equilibrated
in a metastable basin; rather, their long lifetimes reflect the
drastic slowing down of all timescales as the temperature is reduced.

From figure~\ref{fig:nuc}, we conjecture that the nucleation time has
the form
\begin{equation}
\tau_\mathrm{nuc} = \gamma e^{2D/T_c} \frac{e^{4D/T}}{1-(T/T_c)}
\label{equ:t_nuc}
\end{equation}
where $\gamma$ is a (microscopic) rate that depends only weakly on $T$
and $T_c$.  Physically, the nucleation rate is suppressed by a
Boltzmann factor that arises from the activated dynamics, and by a
factor of $\Delta \mu/T\simeq(T_c/T)-1$, where $\Delta\mu$ is the free
energy difference between ordered and disordered phases.

In other words, behaviour near $T_c$ is characterised by a separation
of the nucleation time from the relaxation time, $\tau$. The minimum
in the nucleation time occurs at
\begin{equation}T_x = T_c [ 1 - (T_c/4D) +
\mathcal{O}(T_c/D)^2 ]
\end{equation}
The relaxation time in the paramagnet given approximately by $\tau =
0.14 e^{6D/T}$, so at the minimum we have $\tau^* \simeq 0.14
e^{(6D/T_c)+(3/2)}$ and $\tau_\mathrm{nuc}^* \simeq 0.17 (4D/T_c)
e^{(6D/T_c)+1}$. The result is that $\tau_\mathrm{nuc}^* > \tau^*$ as
long as $T_c < T_o \sim D$, which is the regime of interest in this
section. The two timescales are well separated for all temperatures
between $T_x$ and $T_c$.

This separation results from the small amount of free energy that is
released on ordering.  States in which these times are well-separated
are `supercooled', in the sense that the degrees of freedom associated
with the relaxation time appear to equilibrate in a state that is
known to be unstable at long times.

At very low temperatures, we see that $\tau_\mathrm{nuc}$ will become
smaller than the extrapolated relaxation time $\tau_{J=0} \sim
e^{6D/T}$. The result is that the physical relaxation time at low
temperatures is smaller than $e^{6D/T}$. The dynamics of the aging
state are faster than those of a similar state with $J=0$.

These results may be interpreted in the picture of the model as a
combination of a zero temperature dynamical fixed point and a finite
temperature thermodynamic singularity.  The spin relaxation time is
controlled by the activated dynamics associated with the zero
temperature fixed point. It is large compared to microscopic
timescales, but small compared to the lifetime of the supercooled
state.  That lifetime is very long near the thermodynamic transition:
the slowing down is due to the small free energy difference between
paramagnetic and ferromagnetic states. We comment here that `soft
modes' at large lengthscales are not relevant to the behaviour
observed in simulations, due to the narrowness of the critical region.

While supercooled states are familiar in systems with first order
phase transitions, they are not usually observed near second order
transitions, such as the one discussed in this work.  In first order
systems, the nucleation time may be predicted by thermodynamic
arguments. The free energy of a ($d$-dimensional) droplet of ordered
phase in a disordered background is approximately $\sigma
r^{d-1}-\Delta\mu r^d$ where $r$ is the linear size of the droplet,
$\sigma$ the surface tension and $\Delta\mu$ is the free energy
difference between the two phases. Thus, nucleation requires the
formation of a droplet of linear size $r^*\sim \sigma/\Delta\mu$, with
an associated free energy barrier proportional to $\Delta\mu^{-1}$ (in
two dimensions). The nucleation rate is therefore suppressed by a
factor $e^{-\sigma^2/T\Delta\mu}$. This is exponential suppression of
nucleation.

From (\ref{equ:t_nuc}), we see that the SEV model has linear
suppression: the nucleation rate is proportional to $\Delta\mu$. This
second order system has weaker suppression than that predicted for
first order systems.  Since the transition in the SEV model is second
order, there are processes by which the bulk of the system may be
continuously changed from paramagnet to ferromagnet, without a large
free energy barrier. These processes are slow because they require
co-operative motion of many spins, but the exponential slowdown that
would result from a high energy intermediate state is not observed.
While the phenomenology of the SEV model resembles that of first order
systems, the lifetimes of supercooled states tend to be shorter than
those in systems with diverging free energy barriers near $T_c$.  In
this respect, the SEV model is an imperfect model of a
glass-former. However, we argue that the free energy barrier between
ordered and disordered states in first order systems should mean that
the supercooled states are less affected by the critical point than
those of the SEV model. Thus, if the thermodynamic singularity is
largely irrelevant in this model, then we expect it to be even less
relevant in similar first order systems.

\subsection{Rate equation approach and aging behaviour}
\label{sec:rate_sub}

In order to understand the results of the previous section, we give a
brief discussion of the aging behaviour of the system. We parameterise
this behaviour in terms of mean-field rate equations for the
observables $n_p$ and $n_b$. This will provide further evidence that
the supercooled states are characterised by fast dynamics for the
concentration of excited plaquettes, $n_p$, combined with much slower
dynamics for their spatial ordering (measured by $n_b$).

\begin{figure} \begin{center}
\epsfig{file=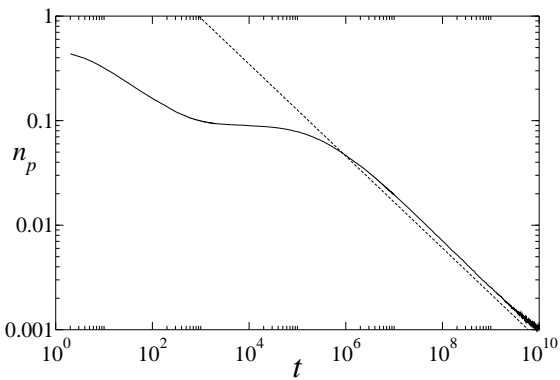,width=0.96\columnwidth}
\end{center}
\caption{Plot showing aging of the paramagnet. We plot
$n_p$ as a function of time after a quench from
infinite temperature to $T=0.286 D$, with $J=0$. The dashed line is
a guide to the eye, corresponding to a decay proportional to $t^{-0.45}$.
}
\label{fig:plaq_age}
\end{figure}

In paramagnetic states with $T<D$, the equilibrium value of
$n_b$ is $0.5$ and that of $n_p$ is $c\simeq e^{-2D/T}$. Aging towards
equilibrium occurs at $n_b=0.5$, with $n_p \sim t^{-0.45}$ 
(a sample trace is shown in figure~\ref{fig:plaq_age}, but the observed
exponent is
independent of temperature, as long we work between $T_c$ and $T_o$).
The rate is limited by the slow diffusion of excited plaquettes 
(there is no simple diffusive process for isolated excitations).

\begin{figure} \begin{center}
\epsfig{file=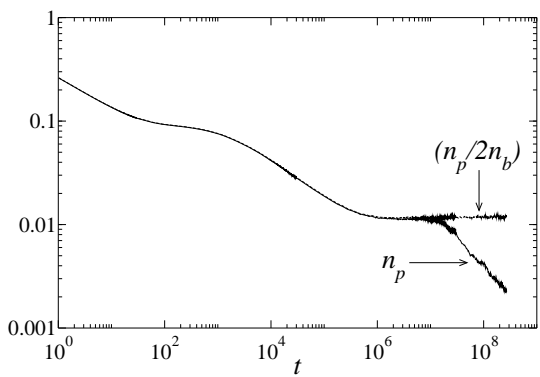,width=0.95\columnwidth}
\end{center}
\caption{Aging of the system at a temperature $T=0.995 T_c$. We have
$T_c=0.451D$ so that $T>T_x$. We plot $n_p$ and $n_p/2n_b$ after a
quench from infinite temperature.  Both quantities are equal until the
supercooled state becomes unstable. At these long times, $n_p$
decreases, but the ratio $n_p/2n_b$ remains constant with a value
approximately equal to $c=e^{-2D/T}$.  }
\label{fig:sc_age}
\end{figure}

There are two regimes for the aging towards the ferromagnetic
state. We define $T_x$ as the temperature at which the nucleation time
is minimal (recall figure~\ref{fig:nuc}).  As shown in
figure~\ref{fig:energy_q}, after quenching to $T$ near $T_c$
($T>T_x$), the system appears to equilibrate at $n_p=c$ (with
$n_b=0.5$), before aging towards the ferromagnetic state.  In
figure~\ref{fig:sc_age} we show a similar scenario, but continuing to
slightly longer times. At these long times, the system ages with the
ratio $(n_p/2n_b)\simeq c$.  The decrease of $n_b$ happens on the long
timescale given by $\tau_\mathrm{nuc}$. However, the plaquette
concentration, $n_p$, is reacting on a much faster timescale. The
interpretation is that $n_p$ has stabilised in an environment set by
the current value of $n_b$. It seems that the stable value of $n_p$ is
around $2cn_b$.

\begin{figure} \begin{center}
\epsfig{file=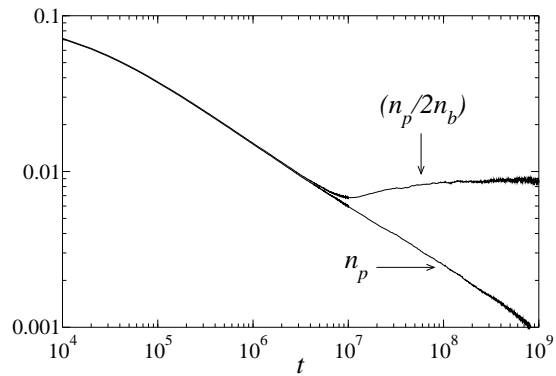,width=0.95\columnwidth}
\end{center}
\caption{Aging at $T=0.364D$, with $T_c=0.444D$, so that $T<T_x$. 
The ratio $n_p/2n_b$ deviates from $n_p$ at a value around
$e^{-2D/T_c}$, so the system does not reach the supercooled state 
with $n_p=c=e^{-2D/T}$.} 
\label{fig:slow_age}
\end{figure}

If instead we quench to a temperature below $T_x$, there is no
apparent equilibration at $n_p\simeq c$. Instead, the effect of finite
$J$ becomes apparent when $n_p\simeq e^{-2D/T_c}>c$. There is a
transient effect as $n_b$ begins to change, but the long time
behaviour again has a constant ratio $(n_p/2n_b)\simeq 0.8e^{-2D/T_c}$
(see figure~\ref{fig:slow_age}). Simulations indicate that the constant
value of $(n_p/2n_b)$ seems to depend on $T_c$ as shown, and to vary
only weakly with $T$. Further, the exponent associated with the aging
appears to be similar to that in the paramagnetic state (around
$0.45$). The natural interpretation is that the aging of these states
is controlled in the same way as the aging of the paramagnet: by the
slow diffusion of excited plaquettes.  The dynamics of $n_b$ are now
faster than those of $n_p$, and the spatial ordering of the excited
plaquettes is responding faster than the plaquette density.

This description of the aging behaviour is consistent with the
following conjectured rate equations
\begin{eqnarray}
e^{4D/T} \partial_t n_p &=& \lambda_p 
n_p^{2.2} ( 2cn_b - n_p) 
\label{equ:mf_np}
\\
e^{4D/T}e^{2D/T_c} \partial_t n_b &=& \lambda_b  
n_b [1-(T/T_c)] 
( X_b - 2 n_b)
\nonumber\\ \label{equ:mf_nb}
\end{eqnarray}
where $X_b=\mathrm{min}(1,n_p A e^{2D/T_c})$.  The significance of
this quantity is that $n_b$ would equilibrate to $X_b/2$ if the
concentration of plaquettes were constrained to be equal to $n_p$.
The exponent $2.2$ in equation~(\ref{equ:mf_np}) is fixed by the
exponent $0.45$ for the power law decay of the energy in the 
paramagnetic phase (as shown in figure~\ref{fig:plaq_age}). 
The the overall scaling of time with
temperature in that equation is also fixed by the aging of the
paramagnetic states. The scaling of time in (\ref{equ:mf_nb})
is determined by the scaling of the nucleation time in
equation~(\ref{equ:t_nuc}).

The adjustable parameters of the theory are therefore the rates
$\lambda_p$ and $\lambda_b$, and the constant, $A$. 
The two rates 
are microscopic frequencies reflecting the co-operative motion of the
spins required to change $n_p$ or $n_b$. The constant $A$ determines
the ratio of $n_b$ and $n_p$ when the aging is controlled by plaquette
diffusion (as in figure~\ref{fig:slow_age}). The instability of the supercooled 
state means that 
$Ae^{2D/T_c} < e^{2D/T}$: the data of figure~\ref{fig:slow_age} is consistent
with $(1/A)\simeq0.8$, as mentioned above. 

We make no attempt to justify these rate equations on microscopic
grounds. For example, the exponent $2.2$ in equation~(\ref{equ:mf_np})
is a non-trivial decay exponent for an annihilation-diffusion problem
in which diffusion of single excited plaquettes is suppressed by the
kinetic constraint. We imagine `integrating out' all the microscopic
degrees of freedom: the effect of the complicated fluctuation effects
appears only in this exponent. However, while their microscopic origin
is unclear, the interesting feature of these equations are (1) the
different temperature scaling of the times associated with the two
degrees of freedom, and (2), the presence of points at which one
degree of freedom is not changing. The first feature leads to a
separation of timescales in the problem. When this is combined with
the second feature, the apparent metastability of the supercooled
states becomes possible. In this case, the faster degree of freedom is
$n_p$, which equilibrates at $2cn_b$ on a timescale that is fast
enough that $n_b$ can be considered to be constant. The aging of the
supercooled state then has a timescale determined by the rate equation
for $n_b$. That degree of freedom is trying to reach apparent
equilibration at $n_pAe^{2D/T_c}/2$, but $n_p$ is the faster degree of
freedom, and is moving the target downwards just as fast as $n_b$
decreases towards it. This results in the aging at constant $n_b/n_p$
that is shown in figure~\ref{fig:sc_age}. An exactly analogous process
is taking place in figure~\ref{fig:slow_age}, except that $n_p$ is now
the slow degree of freedom.

The central point of the above argument is that the slow degrees of
freedom set a `target value' for the fast ones, at which the fast
degrees of freedom appear to equilibrate when viewed on the fast
timescale. This is the sense in which the states discussed in the
previous section are `supercooled'. Their lifetime is then set by the
slow degrees of freedom, and this lifetime is much longer than
relaxation timescales in these states.  To understand the correlations
in the aging state, it would be desirable to study the thermodynamics
of the excited plaquettes while working at fixed $n_b$. However, this
is well beyond the scope of this paper.

Note that there is no provision in equations~(\ref{equ:mf_np}) and
(\ref{equ:mf_nb}) for equilibration in a ferromagnetic state. However,
the large internal energy difference between paramagnetic and
ferromagnetic states means that this equilibration is not observed on
timescales that are accessible in simulation.  Therefore,
equations~(\ref{equ:mf_np}) and~(\ref{equ:mf_nb}) appear to be valid
over timescales that are several orders of magnitude longer than the
lifetimes of the supercooled states.

\begin{figure} \begin{center}
\epsfig{file=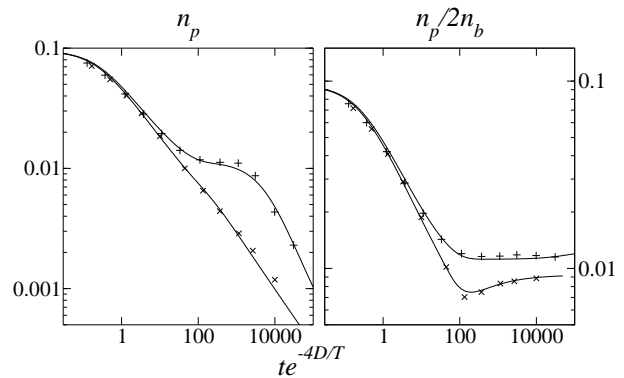,width=0.95\columnwidth}
\end{center}
\caption{Fit to the data of figures~\ref{fig:sc_age} ($+$ symbols)
and~\ref{fig:slow_age} ($\times$ symbols) using mean field rate
equations (\ref{equ:mf_np},\ref{equ:mf_nb}).  Note that the left and
right panels each show data from two runs, showing how the different
regimes are captured by the same rate equations. Parameters are
$(\lambda_p,\lambda_b,A)= (350,17,0.85)$ for the data of
figure~\ref{fig:sc_age}, and $(\lambda_p,\lambda_b,A)= (350,17,1.20)$
for that of figure~\ref{fig:slow_age}.  }
\label{fig:mf_fit}
\end{figure}

It is simple to verify that the qualitative behaviour of
figures~\ref{fig:np_cool} and \ref{fig:plaq_age}-\ref{fig:slow_age}
may be fitted by equations~(\ref{equ:mf_np}) and~(\ref{equ:mf_nb}),
with appropriate values of $(\lambda_p,\lambda_b,A)$. See
figure~\ref{fig:mf_fit}, in which we show reasonable agreement. Note
however, that the onset of nucleation from the supercooled state is
more sudden than that predicted by the mean field equations.  The
initial ordering is slower than the power law predicted by these
simple rate equations.

The requirement of different parameters to fit the different
simulations in figure~\ref{fig:mf_fit} show the rather simplistic
assumptions for the temperature scaling in the mean field
equations. That is, the linear suppression of the nucleation rate with
$(T_c-T)$ is valid only near $T_c$, necessitating adjustment of
$\lambda_b$ at smaller temperatures.  We have already commented that
$A$ will be temperature dependent, but that this dependence has a
weaker effect on $X_b$ than the exponential dependence of that
quantity on $T_c$.

The fits of figure~\ref{fig:mf_fit} using these mean field equations
support our interpretation of figure~\ref{fig:nuc} in terms of purely
dynamical effects that do not depend on thermodynamic quantities like
free energy barriers and spinodal points.

\section{Discussion and Conclusions}
\label{sec:conc}

We have shown that the SEV model can be interpreted in terms of a high
temperature state in which excitations are free and point-like, and a
low temperature phase in which these free excitations are confined
into composite objects, with a characteristic size that is smaller
than their spacing. The dynamics of the paramagnet are well described
in a `mobility field' picture similar to the FA model.

Near the transition to the ordered phase, `supercooled' paramagnetic
states have long lifetimes. Two-time correlation functions show
stationarity in these states, and both their dynamics and their
thermodynamics are well-described by the plaquette model. However,
these states have finite lifetimes, beyond which stationarity is lost
and it becomes clear that they are unstable to the ferromagnetic
state. This situation resembles the situation in physical
glass-formers, except that the presence of a first order transition in
those systems means that the lifetimes of the supercooled states
diverge much faster near $T_c$ than in the SEV model.

We have also shown that the mean field rate equations, (\ref{equ:mf_np})
and (\ref{equ:mf_nb}), are a suitable framework for describing 
the out of equilibrium (aging) behaviour of the SEV model.

We end this work with some comments about the significance of these
results in the context of the literature. Comparing with the work of
Cavagna~\emph{et al.}\cite{Cavagna2003}, we note the similarity
between the phenomenology of their model and the SEV model. However,
the exact solution of the eight vertex model, and the understanding of
the dynamics of the paramagnetic state that arises from previous work
on the plaquette model allow us to investigate the behaviour from a
more microscopic viewpoint.

We illustrate this with three points.  Firstly, the internal energy of
the SEV model changes very rapidly around $T_c$.  From simulation
evidence alone, we might (erroneously) conclude that the transition
was first order.  However, we know from theoretical considerations
that the transition is second order.  This knowledge is important when
discussing the possible metastability of supercooled states.
Secondly, the power law suppression of nucleation near $T_c$ in the
SEV model results from the fact that our transition is second
order. For a first order transition we would expect an exponential
suppression. The nature of this suppression in \cite{Cavagna2003} does
not seem to be clear.  And finally, we are able to identify the
minimum in figure~\ref{fig:nuc} as arising not from the crossing of a
spinodal, but rather from a crossover in timescales.

We would also like to point out some similarities between the
paramagnetic phase of the SEV model and the models which are
conjectured to be controlled by the behaviour of entropic droplets
\cite{Xia-Wolynes,Bouchaud2004cm}.  As mentioned above, the invariance
of the Hamiltonian at $J=0$ under flipping a whole row or column of
spins leads to a large (but non-extensive) ground state entropy.  We
have shown that the introduction of $J$ is largely irrelevant above
$T_c$, and in the supercooled states. Therefore, we may interpret the
low temperature paramagnetic states ($T_c<T\ll T_o$) as a `mosaic' of
the many $J=0$ ground states.  The `droplets' are referred to as
`entropic' in this scenario. This name arises because there does not
seem to be an energetic argument for their stability, so it assumed
that they are stabilised by some entropic mechanism.  In the plaquette
model, the borders between `droplets' of each ground state are not one
dimensional as one might expect, but rather arise from point-like
excited plaquettes. This situation was alluded to in the recent paper
of Bouchaud and Biroli\cite{Bouchaud2004cm}. Since there is no
perimeter for a surface tension to couple to, the entropic forces are
strong enough to stabilise the `mosaic' state. This is the situation
in the SEV model at finite $J$, but $T>T_c$.

If we accept the plaquette model as a realisation of the entropic
droplet picture, it is interesting to note that the only fixed point
in that model is at zero temperature. That is, although the entropy
falls rapidly as the temperature is reduced through $T_o$, any
extrapolation that leads to $S=0$ at a finite temperature is not
valid.  Rather, the entropy remains regular at all temperatures (even
those in which the glassy phase is completely unstable to the
crystal).  In fact, $S_\mathrm{glass} > S_\mathrm{crystal}$ all the
way down to $T=0$ (where $S_\mathrm{glass}$ is the entropy associated
with the plaquette model and $S_\mathrm{crystal}$ is the entropy of
the SEV model with the same value of $D$ and at the same
temperature). In this scenario the Kauzmann paradox \cite{Angell} is
seen as arising from an unphysical linear extrapolation of the glassy
entropy.

To conclude, we have shown that the plaquette model limit of the SEV
model describes its behaviour in both stable and supercooled
paramagnetic states.  This model resembles both the mobility field
description of glassy systems (as exemplified by the FA model) and the
entropic droplet picture.  However, there is no finite temperature
fixed point in the theory of the glassy states; thus Kauzmann's
paradox is avoided.  Taken together, these results are further
evidence that theories without thermodynamic singularities at finite
temperature are suitable for describing glassy states
\cite{GCTheory,Whitelam2004,Berthier2003}.

\acknowledgments
 
We thank A.~Cavagna, D.~Chandler and S.~Whitelam for discussions.  We
acknowledge financial support from EPSRC Grants No.\ GR/R83712/01 and
GR/S54074/01, University of Nottingham Grant No.\ FEF 3024.
 
\begin{appendix}

\section{Low and high temperature expansions of the eight vertex model}
\label{app:series}
In this appendix, we consider the series for the free energy given by
Baxter\cite{BaxterBook}:
\begin{equation}
(f/T) = -\log c' - \sum_m \frac{x^m ( (q/x^2)^m - 1)^2 ( 1-x^m)^2}
                               {m(1-q^{2m})(1+x^{m/2})}
\label{equ:f_series}
\end{equation}
where $q$, $x^2$ and $c'$ depend on the model parameters $(D/T)$ and $(J/T)$.
The parameter $z$ in Baxter's calculation is $1$ in the SEV model since the
Hamiltonian is invariant under $90^\circ$ rotations.
The dependence
of $q$ and $x^2$ on the original parameters $(D,J)$ is rather
indirect: the main task of this appendix will be to derive simple
relations between $(q,x^2)$ and $(D/T,J/T)$ in the ferromagnetic
phase.

However, we first consider the contribution of the $\log c'$ term
to the free energy. In the 
paramagnetic phase we have 
\begin{equation}
(f_\mathrm{pm}/T) \sim \log c' = 
  \log\left[ e^{D/T} \cosh^{2J/T} + e^{-D/T} \right]
\end{equation}
which is the result quoted as (\ref{equ:f_pm}), and may be used
to calculate properties of the paramagnetic phase.
However, in the 
ferromagnetic phase we have $c'=e^{(D+2J)/T}$, so taking only the leading
term leads to the trivial result: $n_p=n_b=0$. There are no 
excitations in the ferromagnetic phase at this order. We must therefore
estimate the parameters $x$ and $q$ in this phase.

The prescription for calculating these parameters is given by
Baxter\cite{BaxterBook}, but we give a brief review. The ratios
$D/T$ and $J/T$ are used to calculate four parameters $(a,b,c,d)$. 
The partition function is symmetric under interchange of the
four quantities $(a\pm b,c\pm d)$.
We may therefore map the parameters into
the principal regime, to give $(a',b',c',d')$ 
satisfying $c'+d'>c'-d'\geq a'+b'>a'-b'\geq0$. These parameters are used
to calculate $\Delta = (-(c'^2+d'^2)+(a'^2+b'^2))/(2c'd'+2a'b')$ 
and $\gamma=(c'd'/a'b')$. 

In the
ferromagnetic phase we have $\gamma = \exp(4D/T)$. 
We work at $T_c<D$ so $\gamma$ is a very 
large number, and we arrive at
\begin{equation}
\Delta = \cosh(4J/T) [ 1 - \gamma^{-1}
(1+\frac{1}{\cosh(4J/T)}) + \mathcal{O}(\gamma^{-2})]
\end{equation}
The next step is to calculate the parameter $k$, from:
\begin{equation}
\gamma(1+k^2) = \Delta^2(1+\gamma)^2 - (1 + \gamma^2)
\end{equation}
and the result is
\begin{equation}
k = \frac{e^{-4D/T}}{2 \sinh^2(2J/T) [ 1 + \cosh(2J/T) ]} +
\mathcal{O}(\gamma^{-2})
\label{equ:k_approx}
\end{equation}
We have $0<k<1$, and $k$ is the elliptic modulus with nome $q$. That is,
\begin{equation}
-\log q \int_0^1 \frac{\mathrm{d}t }{\sqrt{1-t^2}\sqrt{1-k^2
t^2}}
= \pi \int_0^\infty \frac{\mathrm{d}t}
{\sqrt{1+t^2}\sqrt{k^2+t^2}}
\label{equ:q_def}
\end{equation}
For the series of (\ref{equ:f_series}) to converge quickly, we 
require $q<1$: in that
case (\ref{equ:q_def}) reduces to $q\simeq (k^2/16)$, and therefore we have 
the approximate relation 
\begin{equation}
q \simeq \frac{e^{-8D/T}}{128 \sinh^4(2J/T)} 
\end{equation}
whose condition for validity is that $k$ be small, which requires that
$\sinh(2J/T)\ll e^{-2D/T}$. This is the condition that we are 
well inside the ferromagnetic phase, as is clear from (\ref{equ:def_tc}).

In order to evaluate the terms in (\ref{equ:f_series}), 
we also require an approximate form for $x$. The definition of $x$ is
\begin{equation}
\log\left(\frac{q}{x^2}\right)
 \int_0^1 \frac{\mathrm{d}t }{\sqrt{1-t^2}\sqrt{1-k^2
t^2}}
=  \int_{\sqrt{\gamma k}}^\infty \frac{\pi\, \mathrm{d}t}
{\sqrt{1+t^2}\sqrt{k^2+t^2}}
\end{equation}
from which the relation $x^2 < q$ is clear. This integral may be
expanded in a series around $(1/\sqrt{\gamma k})=0$. 
Equation~(\ref{equ:k_approx})
shows that this parameter is small if $J\ll T$. 
The result is that 
\begin{equation}
(q/x^2)-1 \simeq 4 \sinh (2J/T)
\end{equation}
Substituting into (\ref{equ:f_series}), and ignoring all terms with
$m\geq 2$, we arrive at 
\begin{equation}
f_\mathrm{fm} \simeq -(D/T) - (J/T) - \frac{e^{-8D/T}}{16 \sinh^2
(2J/T)}
\end{equation}
which gives the result (\ref{equ:f_fm}),
qualified by the validity condition 
\begin{equation} e^{-2D/T} \ll \sinh(2J/T) \ll 1 
\end{equation}
from which we note that this is not an expansion about $T=0$, but 
rather a useful approximation to the free energy in this part of
the parameter space.

\end{appendix}

\end{document}